\documentclass[aps,pra,preprint]{revtex4-1}
\usepackage{graphicx}  
\usepackage{dcolumn}   
\usepackage{bm}        
\usepackage{amssymb}   
\usepackage{natbib}
\usepackage{amsmath}

\begin{document}

\title{Coherent Phase Control of Internal Conversion in Pyrazine}
\author{Robert J. Gordon}
\affiliation{Department of Chemistry (m/c 111), University of Illinois at
Chicago, Chicago, IL 60680-7061, USA}
\email{rjgordon@uic.edu}
\author{Zhan Hu}
\affiliation{Institute of Atomic and Molecular Physics, Jilin University, Changchun 130021 P.R. China}
\author{Tamar Seideman}
\affiliation{Department of Chemistry, Northwestern University, 2145 Sheridan Road, Evanston, Il 60208, USA}
\author{Sima Singha}
\affiliation{Deerpoint Group, 4339 N Selland Ave., Fresno, CA 93722}
\author{Maxim Sukharev}
\affiliation{Science and Mathematics Faculty, College of Letters and Sciences, Arizona State University, Mesa, AZ 85212, USA}
\author{Youbo Zhao}
\affiliation{Department of Chemistry, University of Illinois at Chicago, Chicago, IL 60680-7061, USA}

\begin{abstract}
\vspace {5mm}
Shaped ultrafast laser pulses were used to study and control the ionization dynamics of electronically excited pyrazine in a pump and probe experiment. For pump pulses created without feedback from the  product signal, the ion growth curve (the parent ion signal  as a function of pump/probe delay) was described quantitatively by the classical rate equations for internal conversion of the $S_2$ and $S_1$ states.  Very different, non-classical behavior was observed when a genetic algorithm (GA) was used to minimize the ion signal at some pre-determined target time, T.  Two qualitatively different control mechanisms were identified for early (T$<1.5$ ps) and late (T$>1.5$ ps) target times. In the former case, the ion signal was largely suppressed for $t<T$, while for $t \gg T$ the ion signal  produced by the GA-optimized pulse and a transform limited (TL) pulse coalesced.  In contrast, for $T>1.5$ ps the ion growth curve followed the classical rate equations for $t<T$, while for $t \gg T$ the quantum yield for the GA-optimized pulse was much smaller than for a TL pulse. We interpret the first type of behavior as an indication that the wave packet produced by the pump laser is localized in a region of the $S_2$ potential energy surface where the vertical ionization energy exceeds the probe photon energy, whereas the second type of behavior may be described by a reduced absorption cross section for $S_0 \rightarrow S_2$ followed by incoherent decay of the excited molecules.

\end{abstract}

\maketitle

\section{Introduction}

The central goal of photochemistry is to use light to understand and control chemical events.
The time-honored approach is to use monochromatic radiation to promote a molecule to an electronically excited state that decays to the desired dissociation or isomerization products. The difficulty  with this  method is that the potential energy landscape of a polyatomic molecule of even modest size is exceedingly complex, and competing processes may intervene as the excited state wave function propagates  from the Franck-Condon (FC) region to the products. A primary example of such intervention is a nonradiative transition (internal conversion, induced by the kinetic energy operator, or intersystem crossing, induced by the spin-orbit operator) to another electronic state, which correlates to different end products. \cite{Yarkony,Yarkony2}

A variety of schemes have accordingly been developed to control the evolution  of excited molecules that exhibit nonradiative transitions, utilizing  the light source as either a ``photonic reagent" or  a ``photonic catalyst." In relatively weak fields, a light pulse may be designed (``shaped") to create a vibrational wave packet that propagates either towards or away from  conical seams connecting different potential energy surfaces (PESs). In this limit, the shaped pulse alters the state population without affecting the PES on which the population evolves. \cite{VR}  In  fields of intermediate strength, the light pulse acts as a catalyst that can distort the energy landscape by either changing the potential energy gradient near the FC region or shifting the location and shape of the conical seams.\cite{Stowlow} In still stronger fields, electrons are removed and subsequent chemistry takes place on a cationic surface. \cite{Chargemigration}

The pyrazine molecule has been used as a benchmark case for  many  theoretical control schemes \cite{Ferretti, Wang, Penfold, Christopher1, Christopher2, Christopher3,BS_3, Thanopulos, Grinev,Sukharev_1, Sukharev_2,Sala} as well as for experimental time-resolved studies of internal conversion (IC). \cite{Horio,Suzuki_1,Radloff}
Reasons for  interest in this molecule are its simple heterocyclic structure ($C_4N_2H_4$),   symmetry ($D_{2h}$), and  small number of valence electrons.  Its pyrimidine isomer serves as a model for three of the DNA bases, and the spectroscopy and photochemistry of pyrazine  have been studied extensively. The first two excited singlet states, $S_1$ and $S_2$, resulting from promotion of either a non-bonding  or a valence  $\pi$ electron to an anti-bonding $\pi^*$ orbital, correspond to diabatic  states with $^1B_{3u}$ and $^1B_{2u}$ symmetries, respectively. \cite{Sobolewski}
These diabatic states are strongly coupled by the $10a$ bending  mode of  $b_{1u}$ symmetry. \cite{Woywod} Because of this strong coupling, the $S_2$ spectrum is very diffuse,\cite{Suzuka,Schneider} and IC occurs in only 22 fs.\cite{Suzuki_1} This ultrafast radiationless transition renders the pyrazine molecule UV-stable, with $S_2 \rightarrow S_1$ occurring much faster than $C-H$ bond-breaking, a property that was critical for the evolution of life on earth. \cite{Domcke_1}

A question of fundamental interest is whether it is possible to create a light pulse that could arrest the rate of IC in pyrazine. A number of theoretical studies in both the weak
\cite{Ferretti, Wang, Penfold, Christopher1, Christopher2, Christopher3,BS_3, Thanopulos, Grinev,Sukharev_1, Sukharev_2}
and intermediate \cite{Sala} intensity regimes answered this question affirmatively, with suppression of IC being achieved for varying periods depending on the control scheme. Inspired by these predictions, we  conducted a pump and probe experiment designed to study the dynamics of electronically excited pyrazine. It is known from time-resolved photoelectron measurements \cite{Oku,Horio} that the $S_1(n^{-1}\pi^*)$ state ionizes primarily to the $D_0(n^{-1})$ cation, whereas $S_2(\pi^{-1}\pi^*)$  ionizes primarily to $D_1(\pi^{-1})$, with the ionization cross section of $S_2$ being $\sim 50\%$ greater than that of $S_1$.\cite{Suzuki_1}  Since we were unable to distinguish experimentally between molecules in the $S_2$ and $S_1$ states, we chose as our objective the more modest goal of controlling the ionization rate, with the expectation that transition from $S_2$ to $S_1$ would produce sharp structure in the ion signal.  We found  that suitably shaped laser pulses  strongly suppressed the parent ion signal for times as long as 1.5 ps, suggesting that the wave packet was localized in a region of the excited state PES having an exceptionally high ionization potential.  A  preliminary report of this finding was published recently, \cite{Pyrazine} and here we extend our initial study and present its findings in  greater detail.

\section{Methods}

The apparatus is shown schematically in Fig. 1. A regeneratively amplified Ti:sapphire laser (Spectra Physics Tsunami oscillator and Spitfire amplifier) peaked around 800 nm with a 24 nm bandwidth produces 35 fs, 3 mJ pulses at a 1 kHz
repetition rate.  A beam splitter directs $85\%$ of the laser energy to an optical parametric amplifier (OPA, Spectra-Physics, TOPAS-C), which generates  30 $\mu$J of 261 nm radiation. The temporal profile of this beam was shaped by an acousto-optic modulator (AOM, Brimrose, FSD3-150-50-240), using a 4f optical configuration and phase-only modulation. The remaining $15\%$ of the energy is converted to a 199 nm probe by passage through a pair of 0.1 mm thick beta barium borate (BBO) crystals.
An optical delay line was used to set the timing between the pump and probe pulses with $<6$ fs precision.

  \begin{figure}[htbp]
   \includegraphics[width=15cm]{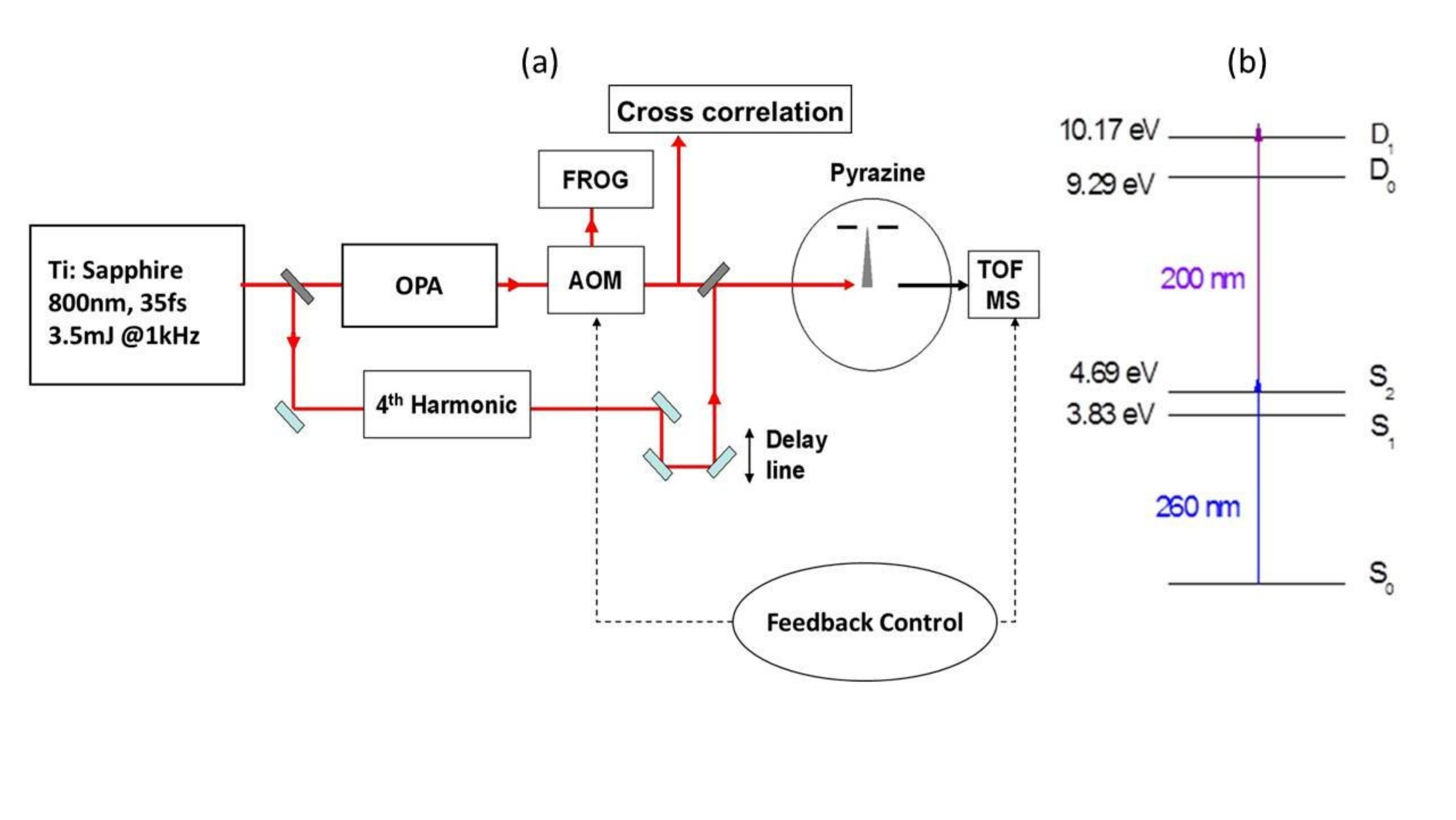}
\caption{Experimental setup. (a) Schematic drawing of the apparatus, including the optical parametric amplifier (OPA),
acousto-optic modulator (AOM), frequency-resolved optical gate (FROG), and time-of-flight mass spectrometer (TOF MS).
(b) Adiabatic energies of the neutral singlet states and the ionic doublet states.  Arrows indicate the energies of the pump and probe photons.}
\end{figure}

A self-diffraction frequency-resolved optical gate (SD-FROG) was used to optimize the alignment of the AOM and facilitate the  generation of transform limited (TL) pulses. A variety of phase functions were applied to the AOM to produce different shaped UV laser pulses. These included a sine phase to generate sequences of evenly spaced TL pulses, \cite{Hornung,Voll,GaAs1,Wollenhaupt} a  V-shaped phase to produce a pulse pair, and a quadratic phase to produce chirped pulses.  The profiles of these pulses were measured with the SD-FROG. The effective time-resolution of the apparatus, taking into account the width of the  pump and probe pulses, was 60 fs.

A continuous beam of pyrazine molecules was produced by expanding the vapor from  crystalline pyrazine (Sigma Aldrich,  $99\%$ purity)  through a 30 $\mu$µm diameter aperture mounted in a differentially pumped vacuum chamber having a base pressure of $2\times 10^{-7}$  Torr. The pump and probe laser beams were focused inside the vacuum chamber using a 200 mm focal length lens mounted on a 3D translation stage. The mass spectrum of the  pyrazine beam generated by the probe laser was measured with a time-of-flight (TOF) mass spectrometer. Z-scans were performed to find the  focal positions of the pump and probe beams and to overlap them with the molecular beam.   The lens was subsequently re-positioned to slightly defocus
the pump beam in order to prevent ionization by the pump alone. The radial spot sizes of the pump and probe pulses at the molecular beam were calculated to be 49.4 $\mu$m and 16.4 $\mu$m, respectively, corresponding to intensities of
$5 \times 10^{10}$ and $3 \times 10^{11}$ W cm$^{-2}$.

In several experiments, a  genetic algorithm (GA) \cite{GA} written in  Visual Basic was used either to minimize or maximize ionization of the pyrazine molecules, utilizing the parent ion signal as a feedback to the pulse shaper. The GA started with a random  distribution of phases comprised of 46 genes. In each evolutionary cycle, the best $50\%$ of the parents were selected for crossover and mutation, and, in addition, the best one of the parents was saved without change. Typically 50-100 generations were required to obtain the optimum pulse shape. The envelopes of the resulting pulses  were measured with a cross-correlator, which used a 0.125 mm thick BBO crystal for sum frequency mixing of the UV probe and the IR fundamental beams.

\section{Results}

Two sets of experiments were performed with the  apparatus described above.  In the first set, various ``simple" pulses (i.e., ones generated without feedback) were used to study the ionization dynamics and calibrate the apparatus, without any attempt at control.  With this information in hand, a second set of experiments was performed with the objective of either maximizing or minimizing the parent ion signal at some predetermined time.

  \begin{figure}[htbp]
   \includegraphics[width=15cm]{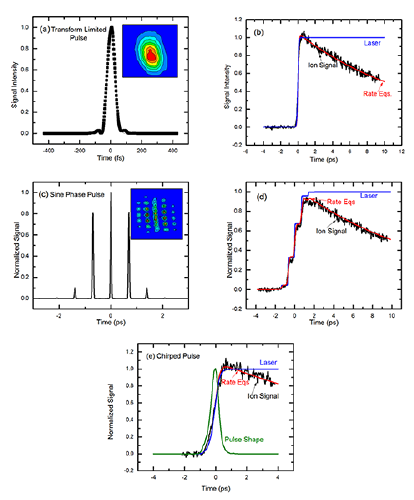}
\caption{Ion signal produced by various simple pulses. (a) Time profile of a transform-limited (TL) pulse, derived from the FROG trace shown in the inset. (b) Experimental ion growth curve (black), calculated ion growth curve (red), and photon growth curve (blue) for the TL pulse.  The peaks of the data and model are normalized to unity, and the decay rate is treated as a least-squares parameter. (c) Time profile of a pulse train produced by a sine phase, derived from the FROG trace shown in the inset. (d) Growth curves for the sine phase pulse.  The scale and decay rate are taken from the  TL signal in panel (b). (e) Growth curves for a positively chirped pulse.}
\end{figure}

The top row of Fig. 2 shows the result obtained with a TL  pulse. Panel (a) displays the pulse envelope determined from the FROG trace shown in the inset, having a full width at half maximum of 55.5 fs.  In panel (b) the ion signal is plotted as a function of pump/probe delay. The smooth red curve is the  solution of the classical rate equations corresponding to the mechanism

\begin{equation}
S_0 \xrightarrow [\text{I(t)$\sigma$}]{} S_2 \xrightarrow [\text{k$_2$}]{}S_1 \xrightarrow [\text{k$_3$}]{} S_0,
\end{equation}
where $I(t)$ is the pump pulse envelope, $\sigma$ is the absorption cross section, $k_2$ is the rate constant for IC from $S_2$ to $S_1$, and $k_3$ is the rate constant for  IC from $S_1$ to $S_0$. The rate equations for this sequence of reactions \cite{Pyrazine}  were integrated numerically, assuming $k_2^{-1} = 22$ fs and treating $\sigma$ and $k_3$ as adjustable parameters. The fitted value of  $k_3^{-1} = 14.3 \pm 0.2$ ps is in good agreement with the literature value of $17.5 \pm 1$ ps. \cite{Radloff}

An experiment was performed to measure possible saturation by the pump pulse.  Using a variable neutral density filter to attenuate the pump, we observed up to a  {$30\%$} deviation from linear dependence of the peak ion signal on pump intensity starting at $1 \times 10^{10}$ W/cm$^2$, corresponding to a pulse energy of 0.05 $\mu$J. When a polarizer and half wave plate were used to attenuate the pulse, the pulse width was stretched to $\sim 100$ fs and the ion signal remained linear throughout.

We next used the fitted values of  $\sigma$ and $k_3$ to analyze the ion signal growth curves for other simple pulses. The middle row of Fig. 2 shows the results for a sequence of five evenly spaced Gaussian pulses generated with a sine phase. Panel (c)  shows the pulse shape determined from the FROG trace, and panel (d) compares the ion signal with the calculated growth curve.  We emphasize that in this and subsequent figures there were no adjustable parameters, inasmuch as  $\sigma$ and $k_3$ were obtained from TL runs done immediately before or after the run with a shaped pulse. Similar results were obtained with chirped pulses, as illustrated in Fig. 2(e). The quantitative agreement of experiment and calculation demonstrates that the the classical rate equations are perfectly adequate for these simple pulses, with quantum effects, if any, exhibited on a time scale too short to be resolved.

A second set of experiments was performed using a GA to  minimize the ion signal at a predetermined time, $T$, equal to 1.0, 1.2, 2.0, 3.0, 4.2, and 5.0 ps. The top row of Fig. 3 shows the result for $T=1$ ps. Panel (a) compares the ion signals from the GA and its companion TL run, where it is evident that very little ionization occurs for $t<T$.  Panel (b) compares the ion growth curve with the rate equation prediction and the photon growth curve (defined as the fraction of photons in the pulse train that have arrived by time $t$). We see that the classical rate equations fail to describe the kinetics for $t<T$; the rate equations predict that the signal should reach $\sim 80\%$ of its maximum value at at $t=T$, whereas the experimental fraction is only $\sim 20\%$.  At long times, however, we find that the TL signal, the GA signal, and the calculated curve all coalesce, indicating that the quantum yield is the same for the TL and GA pulses. Evidently, at early times the excited molecules are transparent at the probe wavelength. The pulse envelope calculated using the GA-optimized phases is superposed on the data in panel (c). This pulse is stretched over a period of approx. 10 ps, with structure contained in two clumps, one  preceding the control time, $T$, and the other following it.  We refer to this collection of results as ``early GA" behavior.

  \begin{figure}[htbp]
   \includegraphics[width=15cm]{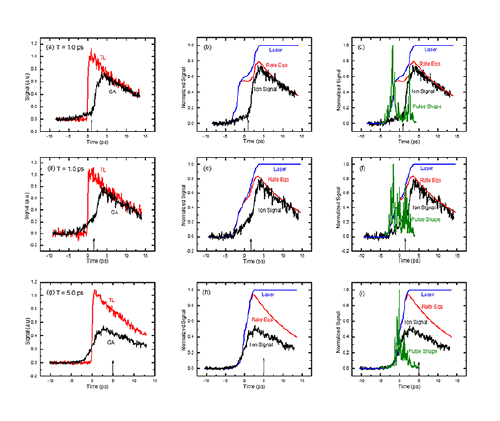}
\caption{Early (top row), intermediate (middle row), and late (bottom row) GA growth curves obtained with   control times of T = 1.0, 1.5, and 5.0 ps, respectively. The left column compares the unscaled ion growth curves obtained with the GA pulses and accompanying TL pulses.   The middle column compares the ion  and photon growth curves with the solution to the rate equations, using the scale parameters and decay rates obtained from the TL control runs. The peak heights of the TL data  were normalized to unity, and GA data and calculated growth curves were scaled accordingly. The right column superposes the GA-optimized pulse envelopes  on the ion growth curves.}
\end{figure}

Similar results were obtained at T=1.2 ps, except that the two lobes of the optimized pulse started to coalesce.  By 1.5 ps, displayed in the middle row of Fig. 3, the ion signal starts to grow at early time, although still showing a sudden rise at $t=T$, and the lobes of the optimized pulse coalesce further. An entirely different type of behavior, referred to as ``late GA", is observed for $T>1.5$ ps. As shown in the bottom row of Fig. 3 for $T=5$ ps, the rate equations track the ion growth curve fairly well at short times,
whereas at long times the ion signal is much smaller than the calculated values or the TL measurement.  In addition, the optimized pulse train has now collapsed into a single lobe.

The relative quantum yield (relative with respect to the TL value) is plotted as a function of the control time in Fig. 4. The distinctly different realms of early and late GA are clearly visible.

  \begin{figure}[htbp]
   \includegraphics[width=15cm]{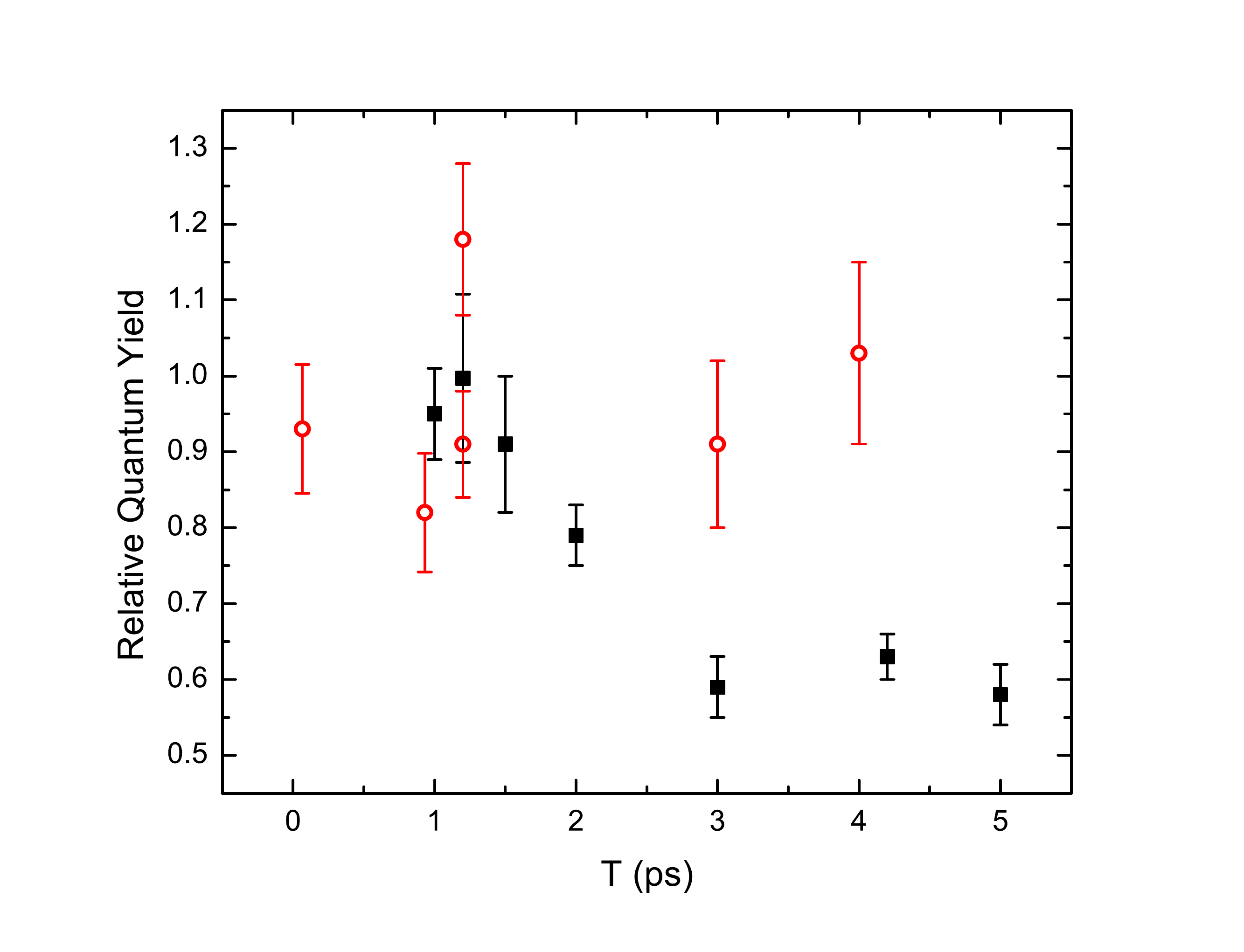}
\caption{Relative quantum yields for the GA set either to  minimize (black squares) or  maximize (red circles) the ion signal at time T.}
\end{figure}

Some discussion of the optimized pulse shapes is in order. The commercial Femtosoft$^\circledR$ program that we used to analyze the FROG traces is unable to invert the complex pulse shapes selected by the GA. Instead,  we used the cross correlator to measure the  envelopes of these pulses. A typical result is shown in Fig. 5a. We found that the AOM is unable to produce the very long pulses called for by the GA, and the measured envelopes are at most 6 ps long.  Within that window, however, the calculated pulse structure follows the experimental one fairly well. Since the  cross-correlator does not fully resolve all the structure in the pulse train, we expect that the actual pulse shape seen by the molecules lies somewhere between the calculated and measured ones.  Fortunately, the growth curves predicted by the rate equations are qualitatively very similar for both the pulse shapes selected by the GA and those measured by the cross-correlator. The calculated growth curves plotted in Fig. 3 used the latter as input to the rate equations.  The various pulses normalized to unit area are  plotted in Fig. 5b to illustrate the dramatic differences between the TL, early GA, and late GA cases. We note that the peak intensities of the GA pulses are well below the  threshold for saturation.

  \begin{figure}[htbp]
   \includegraphics[width=15cm]{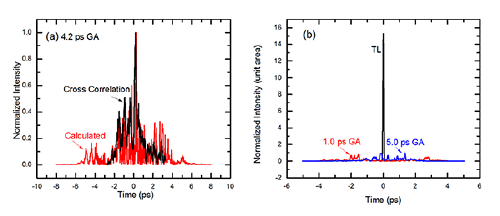}
\caption{Comparison of pulse shapes. Panel (a) compares the pulse envelope calculated from the phase function optimized by the GA programmed to minimize the ion signal at T = 4.2 ps with that measured with the cross correlator. Panel (b) compares the calculated envelopes for a TL pulse and the early and late GA pulses with T = 1.2 and 5 ps, all normalized to unit area.}
\end{figure}

In a final set of experiments we tasked the GA to maximize rather than minimize the ion signal at time $T$.  Two representative examples, one at T=1.2 ps and the other at T=4.2 ps, are shown in Fig. 6. Here we find that the optimum pulse does not alter the relative quantum yield (see Fig. 4) and that the classical rate equations reproduce the data well.

  \begin{figure}[htbp]
   \includegraphics[width=15cm]{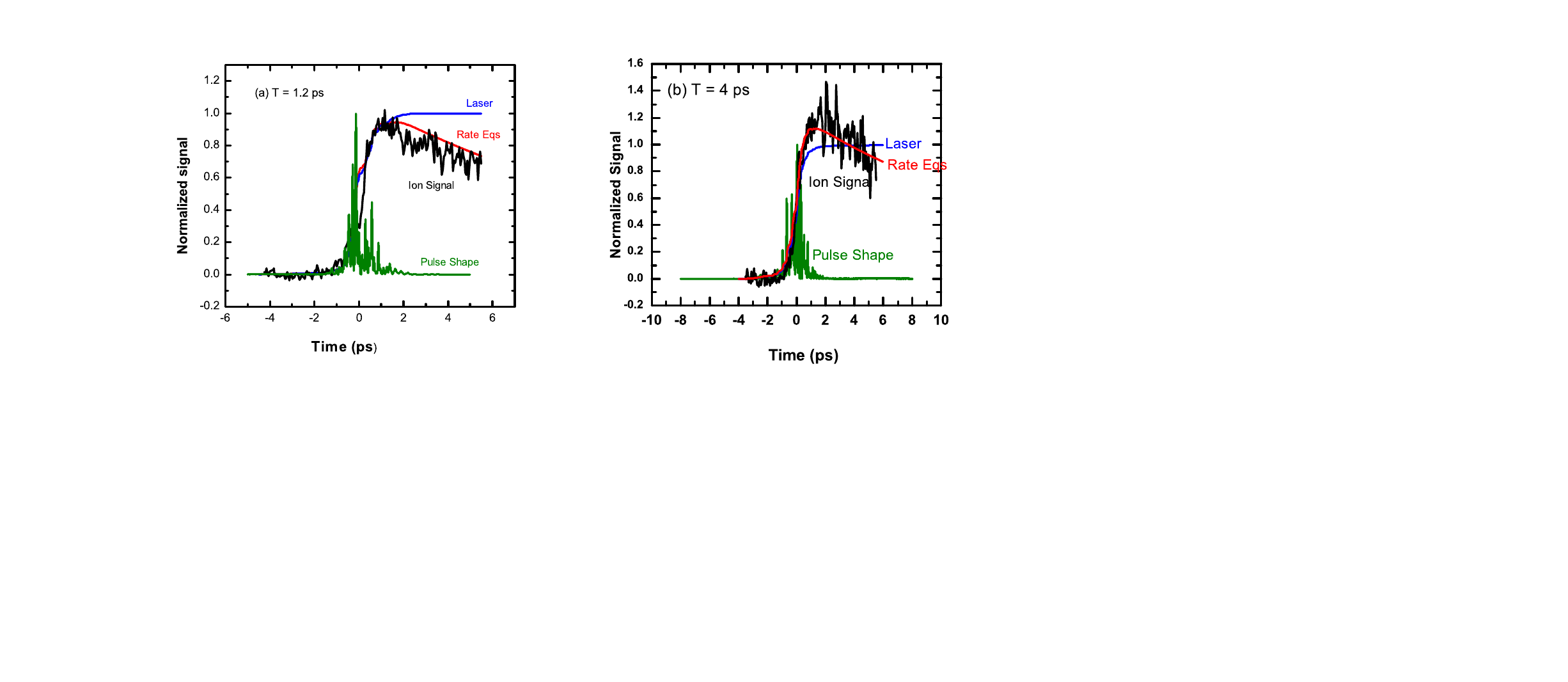}
\caption{Enhancement of the ion signal. The GA was programmed in this case to maximize the ion signal at (a) T = 1.2 ps and (b) T = 4 ps.}
\end{figure}

\section{Discussion}

We have found in this study that in the early GA regime (for target times less than 1.5 ps), the excited pyrazine molecule is transparent to 6.23 eV photons, even though the ionization potential of $S_2$ is only 5.48 eV. We know this because the ion yield at long delay is identical to that of a TL pulse, implying that molecules excited to $S_2$ with an optimally shaped pulse are not immediately ionized by the probe pulse.  In the late GA regime, the reduction in quantum yield may be attributed to a weaker excitation of the $S_2$ bright state resulting from the GA seeking out states with lower absorption cross sections.  When the GA is programmed to maximize the ion signal, there is no change in the quantum yield, and the classical rate equations describe the kinetics well.

A plausible explanation of the early GA transparency is that a vibronic wave packet (WP) produced by the shaped pulse is localized in a region of the $S_2/S_1$ PESs where the vertical ionization potential exceeds the energy of the probe photon. A similar situation was encountered in the transient transparency of ethylene,\cite{Ethylene} where the probe photon energy exceeded the adiabatic ionization potential  by  4.9 eV.  The question we must therefore address is how such a WP could be localized.

A recent theoretical study in the intermediate intensity regime \cite{Sala} found that a non-resonant dynamic Stark effect can shift the conical intersection (CoIn) away from the FC point, thereby suppressing IC of pyrazine while the field is on. In the present experiment, however, the intensity of the control pulse is well below that required for significant Stark modification of the potential energy, and the transparency lasts long after the pulse is over.

There have been numerous theoretical studies in the weak field regime demonstrating coherent control of IC in pyrazine. In the first of these, Ferretti et al. \cite{Ferretti} showed that it is possible to accelerate the rate of IC with a pair of phase-locked $\delta$-function pulses .  Although their model Hamiltonian  was inaccurate, the study demonstrated that a pair of  replica WPs could interfere coherently in the vicinity of a CoIn, thereby modulating the IC rate. Using optimal control theory (OCT) and a 3- or $4$-mode  Hamiltonian, Wang \emph {et al.} \cite{Wang} maximized the population of either  $S_1$ or $S_2$  and were able to stabilize the $S_2$ population for as long as 30 fs.  Using local control theory and a $3$-mode model, Penfold et al.  \cite{Penfold} were able to increase, but not decrease, the IC rate of of pyrazine.

A further advance was made in refs. \onlinecite{Sukharev_1} and \onlinecite{Sukharev_2}, where certain  eigenstates of the coupled Hamiltonian were found to be localized on the $^1B_{3u}/S_2$ state, a quantum phenomenon that was shown to correlate with  scars of periodic orbits. \cite{Heller_1,Heller_2}  These  eigenstates have very large projections on single bright states, endowing them with exceptionally long ($>10$ ns) lifetimes.  Optical pulses that could populate these  long-lived states were constructed using OCT. These unique states were found for an approximate Hamiltonian having 2 or 3 vibrational modes. This study focused on exploring a general phenomenon, and its treatment of the potential energy of pyrazine was highly simplified. In a related approach,  Brumer, Shapiro, and coworkers   \cite{Christopher3,Thanopulos,Grinev} expressed the bright states of $S_2$ as linear combinations of eigenstates of the  complete $24$-mode coupled Hamiltonian. By choosing a suitable linear combination of overlapping resonances, they were able to suppress  $S_2\rightarrow S_1$ internal conversion for periods of $50-100$ fs. Likewise, by exciting suitable resonances, they were able to either enhance or reduce the $S_0 \rightarrow S_2$ absorption. It should be noted that a quantitative  comparison between the calculations and the present experiments is precluded because only two excited diabatic states were included, whereas a recent electronic structure calculation \cite{Robbcomm} using an expanded active orbital space revealed the presence of several dark states in close proximity to $S_2$.

It is also useful to approach the problem in a semi-classical framework.  For a molecule with $N$ atoms, the dimensionality of the  seam  of CoIn may be as large as $3N-8.$  Robb and coworkers \cite{Lasorne} devised a procedure for dividing the $3N-8$ modes into three categories: photoactive modes, which reduce the energy difference between the two PESs of interest as the WP  moves away from the FC point, photoinactive modes, which increase the energy difference, and bath or spectator modes, which leave the energy difference unchanged.  By exciting photoactive modes that propel the WP towards  the seam, it is possible to promote IC, whereas excitation of inactive modes may localize the WP on the initially excited surface.   They went on to apply this idea to the well-known channel three problem in benzene \cite{Fielding} and demonstrated that excitation of a photoactive mode is both necessary and sufficient to reach the CoIn and induce non-radiative decay.	Lasorne \emph{et al.} \cite{Robb_1}  showed further how the topography of the CoIn  determines the IC dynamics.  In the case of benzene, a sloped CoIn on the $S_1$ surface was shown to induce radiationless decay back to $S_0$, whereas a peaked CoIn was identified   in the  channel leading to the benzvalene isomer. It was therefore possible to select the product channel by giving  the initial  WP a  momentum component along a specified direction. \cite{Lasorne}
Using the  \emph{ab initio} multiple spawning  method, Thompson and Martinez showed  how  the fluorescence decay rate of benzene depends on the WP trajectory, with part of the population trapped in a minimum in the $S_1$ surface before reaching the seam. \cite{Thompson}

There is a large literature on vibrational control of the reactivity of electronically excited molecules, including, but not restricted, to vibrationally adiabatic reactions.  Of particular interest here are studies that demonstrate the control of  radiationless transitions by shaping the WP in the FC region. In one such study, Geppert and Vivie-Riedle \cite{VR}  showed that it is possible to control a reaction on an excited PES by preventing the WP from reaching the CoIn.  This strategy is effective because often the CoIn lies in a dark region of the PES, and by imparting momentum to the WP it is possible to steer it away from the seam.  The momentum of the WP may be controlled by imparting a relative phase to its real and imaginary parts.  Without actually constructing the requisite laser pulse shapes, they applied  this strategy to the isomerization of cyclohexadiene using OCT.  The excitation pulse was found to induce a pump-dump-pump cycle, which prolonged the lifetime of the WP in the excited state while directing its momentum away from the CoIn.

In a theoretical study that included the explicit shape of the laser pulse, Mitri\'{c} and coworkers \cite{Mitric}  used a closed loop strategy to suppress IC of adenine in water. They used a library of three-parameter sine phase functions  to shape the laser pulse, along with on-the-fly classical dynamics and  Tully's surface-hopping  procedure. \cite{Tully} They found that with a TL pulse the $S_2/S_3$ states decayed to $S_1$ in $\sim 50$ fs and that $S_1$ decayed to $S_0$ in 475 fs.  The shaped pulse continually repumped the $S_2/S_3$ states for the duration of the pump train ($\sim 1$ ps), so that the $S_1$ lifetime was doubled. Once the pulse train ended, the system returned to normal dynamics.

In another theoretical study using shaped WPs (though without specifying the shape of the driving laser pulse), Robb and coworkers \cite{Robb_2} controlled the isomerization of a  cyanine dye by steering a WP towards different locations of the seam. Steering of the WP was achieved by imparting to it momentum along specific normal coordinates. The strategy was to excite either the  torsional normal modes or a planar stretching mode to steer the WP towards one or another CoIn, resulting in either \emph {cis} or \emph {trans} products.  They found that \emph {trans} $\rightarrow$ \emph {cis} isomerization  occurs unless it is specifically prevented. This was accomplished by decreasing the momentum in the skeletal deformation modes  to allow prompt passage of the WP through a CoIn at large twist angles. Yartsev and coworkers \cite{Yartsev_1,Yartsev_2} reported an experimental implementation of the Robb scheme for controlling the isomerization of a cyanine dye in aqueous solution.  Hoki and Brumer \cite{Hoki} argued, however, that isomerization in solution is an incoherent process dominated by dissipative interactions, and  Yartsev's results may be reproduced quantitatively with a one-dimensional treatment of a single torsional mode.

The success of the classical rate equations for TL and other simple pulses as well as for the pulses found by the GA programmed to maximize the ion signal  is indicative of an incoherent mechanism, whereas the failure of the rate equations for the early GA suggest that quantum coherent processes are important in that case. The complex structure of the latter  is suggestive of  a multi-step sequence of up and down transitions, similar to that found in ref. \onlinecite{Sukharev_2} using OCT, although one must be cautious in making such a comparison because of the restricted Hamiltonian used in their calculation. As shown by Han and Shapiro,\cite{Han} a linear intensity dependence of the signal is not inconsistent with a nonlinear, multi-step mechanism. The observed  transparency for times as long as 1.5 ps is consistent with the calculated coherence time of $\sim 0.5$ ps. \cite{Schneider1,Schneider2}  The overlapping resonance mechanism \cite{Christopher1,Shapiro,Grinev2} of Brumer, Shapiro, and coworkers predicted suppression of IC for up to 100 fs, although longer control periods may be possible by this mechanism.

It is difficult at this point to specify which of the above-mentioned mechanisms  explains our data, especially because a full calculation of the  landscape of the excited PESs of pyrazine is not yet available. In particular, additional calculations of  electronic structure and the reaction dynamics are needed to determine whether the control mechanism is primarily electronic in nature, as would be the case if the FC point lies close to the conical seam, or if the mechanism is dominated by nuclear motion, as envisaged in many of the theoretical studies cited above.

\section{Acknowledgements}

The authors wish to thank Drs. Thomas Weinacht and Marija Kotur for their assistance in setting up the AOM and Drs. Paul Brumer, Wolfgang Domcke, Mike Robb, and Moshe Shapiro for fruitful discussions. This work was supported by the National Science Foundation (CHE-0848198) and by the National Science Foundation of China (10774056 and 10974070).

\pagebreak

\bibliographystyle{phaip}

\end{document}